\documentclass[conference]{IEEEtran}
\IEEEoverridecommandlockouts

\usepackage{cite}
\usepackage{amsmath,amssymb,amsfonts}
\usepackage[ruled,vlined]{algorithm2e}
\usepackage{graphicx}
\usepackage{tikz}
\usetikzlibrary{patterns, decorations.pathreplacing, calc}
\usetikzlibrary{shapes.geometric, shapes.misc}
\usepackage{optidef}
\usetikzlibrary{matrix}
\usetikzlibrary{positioning}
\usepackage{textcomp}
\usepackage{xcolor}
\usepackage{amsthm}
\usepackage{pgfplots}
\usepgfplotslibrary{units}
\usepgfplotslibrary{statistics}
\pgfplotsset{compat=1.3}
\usepackage{makecell}
\usepackage{booktabs}
\usepackage{xcolor}
\usetikzlibrary{calc}
\usepackage{enumitem}
\usepackage{pgfplots}
\pgfplotsset{compat=1.18}

\usepackage{microtype}

\DeclareMathAlphabet\mathbfcal{OMS}{cmsy}{b}{n}
\newcommand{\by}{\boldsymbol{y}}

\newcommand{\bx}{\boldsymbol{x}}

\newcommand{\bn}{\boldsymbol{n}}

\newcommand{\bh}{\boldsymbol{h}}

\begin{document}
\bstctlcite{IEEEexample:BSTcontrol}
%
% paper title
% Titles are generally capitalized except for words such as a, an, and, as,
% at, but, by, for, in, nor, of, on, or, the, to and up, which are usually
% not capitalized unless they are the first or last word of the title.
% Linebreaks \\ can be used within to get better formatting as desired.
% Do not put math or special symbols in the title.

\title{Resource Allocation for Cloud Radar Networks with Communication Constraints}

\author{Christian Eckrich,
        Abdelhak M. Zoubir,
        and Vahid Jamali% <-this % stops a space
\thanks{This work has been co-funded by the LOEWE
initiative (Hesse, Germany) within the
emergenCITY center
[LOEWE/1/12/519/03/05.001(0016)/72], and in part by the German Federal Ministry for Research, Technology and Space (BMFTR) under the program of "Souverän. Digital. Vernetzt." joint project Open6GHub plus (Project-ID 16KIS2407).}% <-this % stops a space
}

\maketitle

% As a general rule, do not put math, special symbols or citations
% in the abstract
\begin{abstract}
Distributed radar sensing exploits spatial diversity to resolve occlusions and improve estimation accuracy. Realizing these gains, however, relies on the transmission of high-dimensional radar data to a Fusion Center (FC). This imposes significant demands on the wireless network, especially in dense, dynamic, and interference-prone environments like factories, where resilience and latency are critical. This paper studies the resource allocation problem in a capacity-constrained two-hop cloud radar network. We propose a buffered access protocol where sensors perform local spectral windowing to reduce data rates before transmitting measurements to the FC via intermediate Edge Servers (ESs). The resource allocation is formulated as a mixed-integer optimization problem aimed at minimizing the aggregate Cramér-Rao Lower Bound (CRLB) of the target parameters subject to fronthaul and backhaul capacity constraints. We develop an iterative solution algorithm based on Big-M formulation and Successive Convex Approximation (SCA). The proposed framework efficiently identifies the most informative sensor subsets and optimizes time-frequency resource assignments, ensuring high-fidelity sensing within stringent communication budgets.
\end{abstract}

% no keywords

% For peer review papers, you can put extra information on the cover
% page as needed:
% \ifCLASSOPTIONpeerreview
% \begin{center} \bfseries EDICS Category: 3-BBND \end{center}
% \fi
%
% For peerreview papers, this IEEEtran command inserts a page break and
% creates the second title. It will be ignored for other modes.
\IEEEpeerreviewmaketitle

\section{Introduction}
Radar sensors have become smaller and cheaper, enabling dense deployments in smart infrastructure and industrial environments \cite{Kong_mmWave_2025, Wang_Multi-Modal_2025}. Radar is privacy-preserving and remains reliable in poor visibility and adverse weather, making it a strong complement to other sensing modalities.

In many scenarios, a single radar view is insufficient because important target parameters may be ambiguous or hidden from one aspect angle, for example due to occlusions. 
Networks of multiple spatially distributed radars provide complementary viewpoints \cite{javadi_radar_2020,liang_design_2011}, improve coverage and estimation accuracy, and increase resilience to sensor failures and local blockages \cite{Nanzer_Distributed_2021,canil_oracle_2024}.
These benefits are only realized when measurements are fused jointly rather than estimated locally at each sensor \cite{sun_multi-sensor_2017,chalise_enhancing_2023}. Consequently, the relevant radar measurements must be forwarded to a processing unit, creating a performance bottleneck when communication resources are limited.
The network must prioritize transmissions from sensors that provide the most informative views of the target \cite{godrich_power_2011,yi_resource_2020, sun_robust_2024,yuan_decentralized_2024}. 

The main contributions of this paper are summarized as follows:
\begin{itemize}[leftmargin=*]
    \item We formulate a mixed-integer optimization problem to minimize the aggregate CRLB of target parameters by jointly optimizing sensor selection and time-frequency resource allocation under fronthaul and backhaul capacity constraints.
    \item We introduce a pipelined transmission protocol that uses local spectral windowing to compress radar measurements, enabling efficient data transfer to the Fusion Center via intermediate ESs.
    \item The original problem is a non-convex mixed-integer program. To cope with this issue, we develop an iterative algorithm based on Big-M formulation and successive convex approximation (SCA) with adaptive regularization. We further demonstrate via simulations that the proposed framework significantly outperforms equal-resource baselines, particularly with tight capacity constraints.
\end{itemize}
\vspace{-0.3cm}
\section{System Overview}

\begin{figure}
    \centering
    \resizebox{0.9\linewidth}{!}{
    % Requires: \usetikzlibrary{patterns,positioning,calc}

\begin{tikzpicture}[>=stealth, scale=0.9, font=\small]

% --- Dimensions ---
\def\wChan{1.2}      % Width of Channel Estimation
\def\wSlot{1.6}      % Width of one Time Slot
\def\hBand{1.2}      % Height of one Frequency Band
\def\vGap{0.3}       % Vertical gap between bands
\def\yS{0.5}         % Y start of Sensing
\def\yF{\yS + \hBand + \vGap} % Y start of Fronthaul
\def\yB{\yF + \hBand + \vGap} % Y start of Backhaul
\def\xStart{0.2}     % X start

% --- Colors ---
\definecolor{colS}{RGB}{100,150,255} 
\definecolor{colF}{RGB}{255,150,100} 
\definecolor{colB}{RGB}{150,255,150} 
\definecolor{col4}{RGB}{200,200,100}

% --- Axes ---
\draw[->, thick] (0,0) -- (13.5,0) node[right] {\textbf{Time ($t$)}};
\draw[->, thick] (0,0) -- (0,5.5) node[above] {\textbf{Frequency ($f$)}};

% --- Channel Estimation (Start) ---
\fill[pattern=crosshatch, pattern color=black!40] (\xStart, \yS) rectangle ++(\wChan, \yB + \hBand - \yS);
\draw[thick] (\xStart, \yS) rectangle ++(\wChan, \yB + \hBand - \yS);
% Label for channel est
\pgfmathsetmacro{\yCenter}{\yS + 0.5*(\yB + \hBand - \yS)}
\node[align=center, font=\scriptsize, rotate=90] at (\xStart + 0.5*\wChan, \yCenter) {Channel Est. \& Resource Alloc.};

% --- Define the Grid Start ---
\pgfmathsetmacro{\xGrid}{\xStart + \wChan}

% --- Helper Macro for Filled Data Block ---
% #1: Slot Index (0-based)
% #2: Y Position
% #3: Color
\newcommand{\datablock}[3]{
    \pgfmathsetmacro{\bx}{\xGrid + #1*\wSlot}
    \fill[#3!70] (\bx, #2) rectangle ++(\wSlot, \hBand);
}

% --- Fill Active Blocks & Add Packet Labels ---
% Pipeline 1 (Blue)
\datablock{0}{\yS}{colS}
\datablock{1}{\yF}{colS}
\datablock{2}{\yB}{colS}
\node[font=\footnotesize\bfseries] at (\xGrid + 2.5*\wSlot, \yB + \hBand + 0.3) {Packet 1};

% Pipeline 2 (Orange)
\datablock{1}{\yS}{colF}
\datablock{2}{\yF}{colF}
\datablock{3}{\yB}{colF}
\node[font=\footnotesize\bfseries] at (\xGrid + 3.5*\wSlot, \yB + \hBand + 0.3) {Packet 2};

% Pipeline 3 (Green)
\datablock{2}{\yS}{colB}
\datablock{3}{\yF}{colB}
\datablock{4}{\yB}{colB}
\node[font=\footnotesize\bfseries] at (\xGrid + 4.5*\wSlot, \yB + \hBand + 0.3) {Packet 3};

% --- Draw The Grid (Empty and Filled) ---
\foreach \i in {0,1,2,3,4,5} {
    \pgfmathsetmacro{\bx}{\xGrid + \i*\wSlot}
    
    % Sensing Band Grid (Horizontal Sections)
    \draw[thick] (\bx, \yS) rectangle ++(\wSlot, \hBand);
    \foreach \h in {0.2, 0.4, ..., 1.0} {
        \draw[black!40] (\bx, \yS+\h) -- ++(\wSlot, 0);
    }
    \draw[thick] (\bx, \yS) rectangle ++(\wSlot, \hBand); % Redraw border

    % Fronthaul Band Grid (Time-Freq Split)
    \draw[thick] (\bx, \yF) rectangle ++(\wSlot, \hBand);
    \draw[step=0.2, black!80] (\bx, \yF) grid ++(\wSlot, \hBand);
    \draw[thick] (\bx, \yF) rectangle ++(\wSlot, \hBand); % Redraw border

    % Backhaul Band Grid (Time-Freq Split)
    \draw[thick] (\bx, \yB) rectangle ++(\wSlot, \hBand);
    \draw[step=0.2, black!70] (\bx, \yB) grid ++(\wSlot, \hBand);
    \draw[thick] (\bx, \yB) rectangle ++(\wSlot, \hBand); % Redraw border
}

% --- Channel Estimation (End) ---
\pgfmathsetmacro{\xEnd}{\xGrid + 6*\wSlot}
\fill[pattern=crosshatch, pattern color=black!40] (\xEnd, \yS) rectangle ++(\wChan, \yB + \hBand - \yS);
\draw[thick] (\xEnd, \yS) rectangle ++(\wChan, \yB + \hBand - \yS);
% Label
\pgfmathsetmacro{\yCenter}{\yS + 0.5*(\yB + \hBand - \yS)}
\node[align=center, font=\scriptsize, rotate=90] at (\xEnd + 0.5*\wChan, \yCenter) {Channel Est. \& Resource Alloc.};

% --- Arrows ---
\foreach \i/\col in {0/colS, 1/colF, 2/colB} {
    \pgfmathsetmacro{\currX}{\xGrid + \i*\wSlot + 0.7*\wSlot}
    \pgfmathsetmacro{\nextX}{\xGrid + (\i+1)*\wSlot + 0.3*\wSlot}
    
    % S -> F
    \draw[->, thick, \col!80!black] (\currX, \yS+\hBand) .. controls ++(0.3,0.4) and ++(-0.3,-0.4) .. (\nextX, \yF);
    
    % F -> B
    \draw[->, thick, \col!80!black] (\nextX + 0.4*\wSlot, \yF+\hBand) .. controls ++(0.3,0.4) and ++(-0.3,-0.4) .. (\nextX + 1.0*\wSlot, \yB);
}

% --- Labels ---
\node[left] at (-0.1, \yS + 0.5*\hBand) {\textbf{Sensing}};
\node[left] at (-0.1, \yF + 0.5*\hBand) {\textbf{Fronthaul}};
\node[left] at (-0.1, \yB + 0.5*\hBand) {\textbf{Backhaul}};

% --- Tc Indicator ---
\draw[<->, thick] (\xGrid, -0.3) -- (\xEnd, -0.3) node[midway, below] {$T_{\mathrm{c}}$};

% --- Continuation Dots ---
\node at (\xEnd - 0.5*\wSlot, \yS + 0.5*\hBand) {$\cdots$};
\node at (\xEnd - 0.5*\wSlot, \yF + 0.5*\hBand) {$\cdots$};
\node at (\xEnd - 0.5*\wSlot, \yB + 0.5*\hBand) {$\cdots$};

\end{tikzpicture}}
    \caption{Schematic of the proposed buffered access protocol showing fronthaul and backhaul phases within a frame.}
    \label{fig:Protocol}
\end{figure}
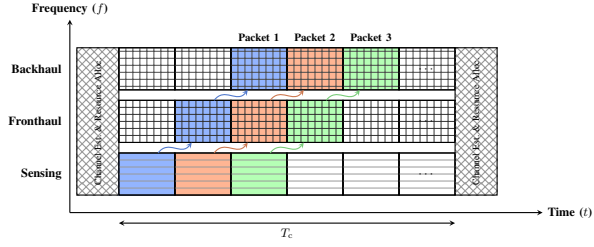

In large-scale deployments, such as factories or campus networks, having every sensor connect directly to a central FC can be impractical. We therefore consider a two-hop cloud radar network with $I$ spatially distributed radar sensors connected to the FC via $J$ intermediate ESs, as illustrated in Fig.~\ref{fig:Protocol}. This architecture improves scalability by having sensors transmit their measurements over capacity-limited fronthaul links to their serving ESs. The ESs then forward the data to the FC over capacity-limited backhaul links for joint parameter estimation.

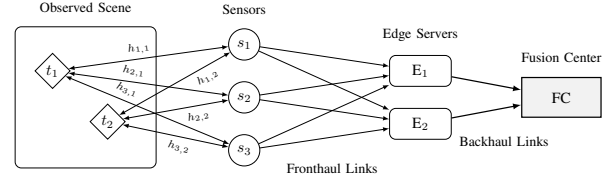
\begin{figure}
    \centering
    \resizebox{0.9\linewidth}{!}{
    \begin{tikzpicture}[
    >=latex,
    node distance = 1.4cm and 2.4cm,
    sensor/.style = {draw, circle, minimum size=7mm},
    target/.style = {draw, diamond, minimum size=8mm, inner sep=1pt},
    scene/.style  = {draw, rectangle, rounded corners, inner sep=8pt},
    es/.style     = {draw, rectangle, rounded corners, minimum width=14mm, minimum height=7mm},
    fc/.style     = {draw, rectangle, thick, minimum width=18mm, minimum height=9mm, fill=gray!10}
]

    % Smaller scene box
    \node[scene, minimum width=3.2cm, minimum height=3.2cm] (scene) at (0,0) {};
    \node[above=0.2cm of scene, font=\small] {Observed Scene};

    % Targets inside scene (slightly adjusted)
    \node[target] at (-0.75,  0.60) (t1) {$t_1$};
    \node[target] at ( 0.50, -0.55) (t2) {$t_2$};

    % Sensors (kept close to scene)
    \node[sensor] at (3.6,  1.2) (s1) {$s_1$};
    \node[sensor] at (3.6,  0.0) (s2) {$s_2$};
    \node[sensor] at (3.6, -1.2) (s3) {$s_3$};

    % Edge servers (moved right to create more space for sensor->ES links)
    \node[es] at (7.6,  0.6) (e1) {E$_1$};
    \node[es] at (7.6, -0.6) (e2) {E$_2$};

    % Fusion center (shifted accordingly)
    \node[fc] at (10.8, 0.0) (fc) {FC};

    % Sensing links (targets <-> sensors)
    \draw[<->] (t1) -- node[pos=0.45, above, sloped, font=\scriptsize] {$h_{1,1}$} (s1);
    \draw[<->] (t1) -- node[pos=0.40, above, sloped, font=\scriptsize] {$h_{2,1}$} (s2);
    \draw[<->] (t1) -- node[pos=0.35, above, sloped, font=\scriptsize] {$h_{3,1}$} (s3);

    \draw[<->] (t2) -- node[pos=0.75, below, sloped, font=\scriptsize] {$h_{1,2}$} (s1);
    \draw[<->] (t2) -- node[pos=0.70, below, sloped, font=\scriptsize] {$h_{2,2}$} (s2);
    \draw[<->] (t2) -- node[pos=0.55, below, sloped, font=\scriptsize] {$h_{3,2}$} (s3);

    % Communication links (sensors -> ES)
    \draw[->] (s1) --  (e1);
    \draw[->] (s2) --  (e1);
    \draw[->] (s3) --  (e1);

    \draw[->] (s1) --  (e2);
    \draw[->] (s2) --  (e2);
    \draw[->] (s3) --  (e2);

    % Communication links (ES -> FC)
    \draw[->, thick] (e1) --  (fc);
    \draw[->, thick] (e2) --  (fc);

    % Layer labels
    \node[above=0.2cm of s1, font=\small] {Sensors};
    \node[above=0.2cm of e1, font=\small] {Edge Servers};
    \node[above=0.2cm of fc, font=\small] {Fusion Center};

    \node[font=\small] at (5.6, -1.6) {Fronthaul Links};
    \node[font=\small] at (9.5, -1.0) {Backhaul Links};

\end{tikzpicture}}
    \caption{Schematic of the cloud radar network showing targets, distributed sensors with sensing channels $h_{i,p}$, and capacity constrained fronthaul/backhaul links to the FC via ESs.}
    \label{fig:Network}
\end{figure}

Communication is scheduled over time-frequency resource blocks, resulting in link-dependent capacities that constrain the achievable sensor and ES flow rates, as illustrated in Fig.~\ref{fig:Network}. In the following, we detail the radar sensing model and the communication protocol.

\subsection{Radar Sensing}
Each radar sensor employs a frequency modulated continuous wave (FMCW) waveform and a uniform linear array with inter-element spacing $d =\lambda/2$. A radar measurement frame consists of $L$ signal chirps of duration $T_c$ and bandwidth $B$. 
Let the complex baseband transmit chirp be
\begin{align}
x_{\text{tx}}(t)=\exp\!\left(j2\pi\left(f_c t+\frac{B}{2T_c}t^2\right)\right),
\end{align}
and consider a single far-field point target. The received signal at antenna element $m\in\{0,\dots,M-1\}$ during chirp $\ell \in\{0,\dots,L-1\}$ can be modeled as
\begin{align}
r_{m,\ell}(t)=\sum_p \alpha_p\,x_{\text{tx}}(t-\tau_{m,p})\,e^{j2\pi f_{D,p} t}+w_{m,\ell}(t),
\end{align}
where $\alpha_p\in\mathbb{C}$ captures path loss and complex reflectivity, $f_{D,p}$ is the Doppler frequency, and $\tau_{m,p}=\tau_0+m\frac{d}{c}\sin\theta_p$ denotes the round-trip delay at antenna element $m$ of the $p$-th target. 

Dechirping is performed by mixing with the conjugate transmit chirp, which yields an intermediate frequency signal that is approximately a complex sinusoid in fast-time, with phase progressions across chirps and antennas. Sampling at $t=nT_s$, where $n\in\{0,\dots,N-1\}$ and $N$ denotes the number of fast-time samples per chirp, results in the discrete data cube
\begin{align}
x[n,\ell,m]
\hspace{-0.1cm}= \hspace{-0.1cm}\sum_p \hspace{-0.1cm}
\alpha_p\,
e^{j2\pi \bigl(f_{r,p} nT_s + f_{D,p} \ell T_c + f_{a,p} m \bigr)}\hspace{-0.1cm}
+ \hspace{-0.05cm}w_{n,\ell,m},
\label{eq:data_cube}
\end{align}
where $f_{r,p}$ denotes the beat frequency, $T_s$ is the fast-time sampling interval, and $T_c$ is the chirp repetition interval. Moreover, $f_{a,p}=\tfrac{d}{\lambda}\sin\theta_p$ is the spatial frequency associated with the angle of arrival $\theta_p$, and $w[n,\ell,m]$ denotes clutter and additive noise.

\subsection{Local Radar Processing}
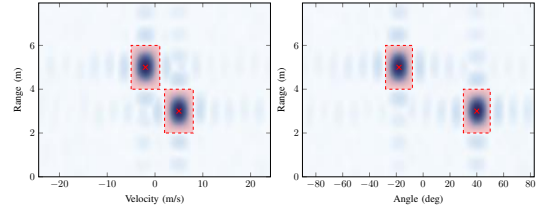
\begin{figure}
    \centering
    \resizebox{0.8\linewidth}{!}{
    % PGFPlots code generated by simulation_kernels.py
% Requires \usepackage{pgfplots}
% \pgfplotsset{compat=1.18}

% --- map_RD ---
\begin{tikzpicture}
    \begin{axis}[
        xlabel={Velocity (m/s)},
        ylabel={Range (m)},
        xmin=-24.351, xmax=24.160,
        ymin=0.000, ymax=7.938,
        scale only axis,
        width=8cm,
        height=6cm,
        enlargelimits=false,
        axis on top,
        % colormap/Blues % Standard PGFPlots library required if using colormesh, here we use graphics
    ]
        \addplot graphics [xmin=-24.351, xmax=24.160, ymin=0.000, ymax=7.938] {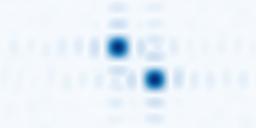};
        \draw[red, dashed, fill=red, fill opacity=0.2] (axis cs: 2.0, 2.0) rectangle (axis cs: 8.0, 4.0);
        \node[cross out, draw=red, inner sep=2pt, thick] at (axis cs: 5.000, 3.000) {};
        \draw[red, dashed, fill=red, fill opacity=0.2] (axis cs: -5.0, 4.0) rectangle (axis cs: 1.0, 6.0);
        \node[cross out, draw=red, inner sep=2pt, thick] at (axis cs: -2.000, 5.000) {};
    \end{axis}
\end{tikzpicture}

% --- map_RA ---
\begin{tikzpicture}
    \begin{axis}[
        xlabel={Angle (deg)},
        ylabel={Range (m)},
        xmin=-90.000, xmax=82.833,
        ymin=0.000, ymax=7.938,
        scale only axis,
        width=8cm,
        height=6cm,
        enlargelimits=false,
        axis on top,
        % colormap/Blues % Standard PGFPlots library required if using colormesh, here we use graphics
    ]
        \addplot graphics [xmin=-90.000, xmax=82.833, ymin=0.000, ymax=7.938] {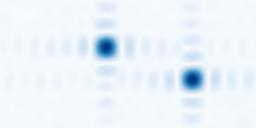};
        \draw[red, dashed, fill=red, fill opacity=0.2] (axis cs: 30.0, 2.0) rectangle (axis cs: 50.0, 4.0);
        \node[cross out, draw=red, inner sep=2pt, thick] at (axis cs: 40.000, 3.000) {};
        \draw[red, dashed, fill=red, fill opacity=0.2] (axis cs: -28.0, 4.0) rectangle (axis cs: -8.0, 6.0);
        \node[cross out, draw=red, inner sep=2pt, thick] at (axis cs: -18.000, 5.000) {};
    \end{axis}
\end{tikzpicture}}
    \caption{Spectral representation of the received radar signal. Red dashed boxes mark local windows around detected targets that are extracted and sent to the FC.}
    \label{fig:SpectralWindows}
\end{figure}

By applying discrete Fourier Transforms (DFTs) along fast-time, slow-time, and spatial dimensions, the data cube is transformed into the spectral domain, where targets appear as peaks in the range-Doppler-angle spectrum. This consequently concentrates the relevant target information into a few dominant coefficients. 
Specifically, the radar maps the data cube $x[n,\ell,m]$ to 
\begin{align}
X[k_r,k_d,k_a]=
\sum_{m,\ell,n}
x[n,\ell,m]\,
e^{-j2\pi\left(\frac{nk_r}{N}+\frac{\ell k_d}{L}+\frac{m k_a}{M}\right)},
\end{align}
where $k_r=0,\dots,N-1$, $k_d=0,\dots,L-1$, and $k_a=0,\dots,M-1$ are the discrete frequency indices in the range, Doppler, and angle dimensions, respectively.
Substituting the signal model \eqref{eq:data_cube} into the DFT definition, the spectral response for a single point target can be expressed as
\begin{align}
    X[k_r,k_d,k_a] \hspace{-0.1cm}= \hspace{-0.1cm}&\sum_p\hspace{-0.1cm} \alpha_p \, D_N(f_{r,p}, k_r) D_L(f_{D,p}, k_d) D_M(f_{a,p}, k_a)\nonumber\\ &+ W[k_r,k_d,k_a],
    \label{eq:spectral_response}
\end{align}
where the Dirichlet kernel is defined as
\begin{align}
    D_N(f, k) = \frac{\sin(\pi N (f - k/N))}{\sin(\pi (f - k/N))} e^{-j\pi(N-1)(f-k/N)}.
\end{align}
Since the DFT is invertible, representing the data cube in the spectral domain is lossless. To reduce the communication load when transmitting data to the FC, we consider a compression strategy based on spectral windowing.
Therefore, each radar first performs local detection using a low threshold. This ensures that all potential target locations are captured, while the FC is responsible for high fidelity estimation and false alarm mitigation. For each candidate detection, a spectral window $\Omega_p$ is extracted, centered at the peak indices $k_p = (k_{r,p}, k_{d,p}, k_{a,p})$, and transmitted to the FC for joint detection and processing. The window is defined as
\begin{align}
\Omega_p
=
\Big\{&
(k_r,k_d,k_a):
|k_r-k_{r,p}|\le\Delta_r,\,\nonumber\\
&|k_d-k_{d,p}|\le\Delta_d,\,
|k_a-k_{a,p}|\le\Delta_a
\Big\},
\end{align}
where $(\Delta_r,\Delta_d,\Delta_a)\in \mathbb{N}$ determine the window size in the range, Doppler, and angle dimensions. Fig.~\ref{fig:SpectralWindows} illustrates the spectral windowing for range-Doppler and range-angle maps. 

This compression induces two error sources:
\paragraph{Truncation Error}
The extraction of the local spectral window $\Omega_p$ introduces a truncation error due to the non-vanishing sidelobes of the Dirichlet kernels in \eqref{eq:spectral_response}. Using the envelope bound $\lvert D_N(f,k)\rvert^2 \le (4\lvert f-k/N\rvert^2)^{-1}$, the discarded energy outside the window can be bounded by a convergent tail sum. In particular, the contribution of the range dimension satisfies $\sum_{\lvert k_r-k_{r,p}\rvert>\Delta_r} \lvert D_N(f_r,k_r)\rvert^2 \le N^2/(2\Delta_r)$, with analogous bounds for Doppler and angle. Since the spectral response in \eqref{eq:spectral_response} factorizes across dimensions, the total truncation error scales as $\mathcal{O}\!\left(\frac{N^2}{\Delta_r}+\frac{L^2}{\Delta_d}+\frac{M^2}{\Delta_a}\right)$, showing that the truncation error decays inversely with the window size in each dimension.

\paragraph{Quantization Error}
In addition, digital transmission of the local spectral coefficients requires quantization. In this work, we assume that the coefficients within $\Omega_p$ are quantized using the same resolution as the radar front-end analog-to-digital converter. Under this assumption, the resulting quantization noise is dominated by the sensor noise floor and does not constitute a performance-limiting factor at the FC. The impact of quantization on sensing performance has been addressed in detail in \cite{Eckrich_Fronthaul_2024}, and is therefore not modeled explicitly in the following.

The discrete spectral coefficients within $\Omega_p$ are stacked into a measurement vector $\by_p \in \mathbb{C}^{|\Omega_p|}$, where 
\begin{align}
    |\Omega_p|=(2\Delta_r+1)(2\Delta_d+1)(2\Delta_a+1),
\end{align}
and transmitted to the associated ESs. Assuming a constant window size for all targets and $q$ bits per complex spectral coefficient, the required number of bits per target is 
\begin{align}
    \beta = q|\Omega_p| + \log_2(NLM),
\end{align}
where the second term accounts for the transmission of the peak index $k_p$. Consequently, the flow rate generated by radar sensor $i$ is given by $R_i = \sum_p v_{i,p} \beta$,
% \begin{align}
%     R_i = \sum_p v_{i,p} \beta,
% \end{align}
where $v_{i,p} \in \{0,1\}$ indicates whether or not sensor $i$ transmits measurements for target $p$.

\subsection{Communication Model}
After an initial channel-estimation stage, sensing and communication are conducted in dedicated bands for sensing, fronthaul, and backhaul. Radar measurements are acquired in batches on the sensing band and buffered locally, while previously acquired batches are relayed upward through the network. The fronthaul and backhaul transmissions are each scheduled over $K$ orthogonal time-frequency resource blocks. Fig.~\ref{fig:Protocol} illustrates the proposed pipelined transmission protocol within one channel coherence interval.

\subsubsection{Fronthaul Phase}
During the fronthaul phase, radar sensors transmit buffered measurements to the associated ESs. Let \(K\) denote the number of resource blocks. The estimated capacity of the link between sensor \(i\) and ES \(j\) on resource block \(k\) is denoted by \(C_{i,j}^{(k)}\). The allocation variable \(a_{i,j}^{(k)} \in \{0,1\}\) indicates if resource block \(k\) is assigned to sensor \(i\) and ES \(j\), subject to the constraint \(\sum_{i,j} a_{i,j}^{(k)} \leq 1\), \( \forall k\).

The effective fronthaul capacity of sensor \(i\) therefore amounts to \(\sum_{j,k} a_{i,j}^{(k)} C_{i,j}^{(k)}\), and must be sufficient to accommodate the flow rate \(R_i\). This leads to the fronthaul constraint
\begin{equation}
    \textbf{C1:} \quad \sum_{j,k} a_{i,j}^{(k)} C_{i,j}^{(k)}\geq R_i. \label{C1}
\end{equation}

\subsubsection{Backhaul Phase}
During the backhaul phase, each ES forwards the aggregated data to the FC using a similar resource allocation scheme. Let \(\bar{C}_j^{(k)}\) denote the estimated capacity of the backhaul link between ES \(j\) and the FC on resource block \(k\). The allocation variable \(b_j^{(k)} \in \{0,1\}\) indicates if resource block \(k\) is assigned to ES \(j\), with the constraint \(\sum_j b_j^{(k)} \leq 1\), \(\forall k\).
Because sensors may split traffic arbitrarily across reachable ESs, we enforce a conservative backhaul bound based on the incoming and outgoing capacities at the ESs, so that 
\begin{equation}
    \textbf{C2:} \quad \sum_{i,k} a_{i,j}^{(k)} C_{i,j}^{(k)} \leq \sum_k b_j^{(k)} \bar{C}_j^{(k)}, \quad \forall j.
\end{equation}

With the introduced constraints for the fronthaul and backhaul phases, the proposed protocol improves resource utilization and ensures causality within the two-hop architecture. Furthermore, it reduces end-to-end latency by transmitting small measurement batches within a single channel coherence interval.

\subsection{Aggregation and Measurement Model}

At the FC, the received spectral windows are used to reconstruct the original full data cube.
As the range, Doppler, and angle parameters for each target are seen differently by each sensor, the FC maps each local measurement vector $\by_{i,p}$ to a common reference frame before fusion. 
The final scene can be estimated using standard techniques such as sparse reconstruction or maximum likelihood estimation.

\section{Problem Formulation}
Our primary objective is to optimize the network resource allocation to maximize the joint estimation accuracy of the target parameters location and velocity at the FC. While these parameters are initially estimated locally to identify the spectral regions of interest, the final achievable precision at the FC is fundamentally limited by the aggregated SNR collected across all sensors. 
Directly optimizing the CRLB for the geometric parameters is intractable, as the bounds depend non-linearly on the unknown target parameters. Instead, we focus on optimizing the estimation of the complex reflectivity $\alpha_p$ for each target, which is directly related to the integrated SNR.

For a detected target $p$, the stacked measurement vector $\by_p$ at the FC is modeled as
\begin{align}
    \by_p &= 
    \begin{bmatrix}
        \by_{1,p}^T & \hdots & \by_{I,p}^T\\
    \end{bmatrix}^T\nonumber\\ &= 
    \begin{bmatrix}
        v_{1,p}\bh_{1,p}^T&\hdots& v_{I,p}\bh_{I,p}^T
    \end{bmatrix}^T
    \alpha_p + \bn_p,
\end{align}
where $\bn_p$ is the aggregate noise. The vectors $\bh_{i,p}, \forall i$ capture the sensing channel for sensor $i$ and target $p$.

Under the assumption of spatially uncorrelated noise with variances $\sigma_i^2$, the Fisher Information (FI) for the complex reflectivity $\alpha_p$ reduces to the scalar sum
\begin{align}
    \text{FI}_p = \sum_{i} v_{i,p} \bh_{i,p}^H \frac{1}{\sigma_i^2} \bh_{i,p}.
\end{align}
The optimization objective is to minimize the sum of the CRLBs across all targets given by
\begin{align}
    \sum_p \text{CRLB}_{\alpha_p}(v_{i,p},\forall i) = \sum_p \left( \text{FI}_p \right)^{-1}.
\end{align}
This objective function drives the network to select sensors that provide the highest SNR for each target. Furthermore, it inherently ensures that every target is observed by at least one sensor, as an unobserved target would result in an infinite CRLB.
The resulting optimization problem can then be formulated as follows. 
\begin{mini!}|s|
    {\hspace{-1.0cm}\substack{\hspace{1.0cm}v_{i,p},a_{i,j}^{(k)}, b_{j}^{(k)}, \\\hspace{1.0cm}\forall i,j,p,k}}{\sum_p \text{CRLB}_{\alpha_p}(v_{i,p},\forall i)}%
    {\label{Eq:P1}}{\textbf{P1:}}
    \addConstraint{\hspace{-0.7cm}\textbf{C1}, \textbf{C2}\nonumber}{}
    %\addConstraint{\hspace{-0.9cm}\textbf{C1:}\;}{\sum_{p}v_{p}r\leq \sum_{i,k}a_{i,j}^{(k)}C_{i,j}^{(k)},\;\forall i \label{eq:SensorFlowCons}}
    %\addConstraint{\hspace{-0.9cm}\textbf{C2:}\;}{\sum_{i,k}a_{i,j}^{(k)}C_{i,j}^{(k)}\leq \sum_{k}b_{j}^{(k)}\bar{C}_j^{(k)},\;\forall j \label{eq:EdgeFlowCons}}
    \addConstraint{\hspace{-0.9cm}\textbf{C3:}\;}{\sum_{i,j}a_{i,j}^{(k)}\leq 1,\;\sum_{j}b_{j}^{(k)}\leq 1,\;\forall k \label{eq:FBHaulCons}}
    \addConstraint{\hspace{-0.9cm}\textbf{C4:}\;}{v_{i,p},a_{i,j}^{(k)}, b_{j}^{(k)}\in \{0,1\},\forall i,j,p,k \label{eq:IntegerCons}}
\end{mini!}
The optimization problem \textbf{P1} is a non-convex Mixed-Integer Linear Program (MILP). The binary variables make the problem NP-hard in general and the non-convex objective further complicates finding a globally optimal solution.
\vspace{-0.2cm}
\subsection{Problem Reformulation}
To obtain a tractable formulation, we first linearize the objective function using Big-M formulation, relax the integer constraints, and apply SCA to promote binary solutions.

\subsubsection{Objective Linearization}
Let $z_{i,p} = \bh_{i,p}^H \frac{1}{\sigma_i^2} \bh_{i,p} \in \mathbb{R}^+$ denote the contribution of sensor $i$ to the FI of target $p$. The inverse FI of target $p$ can then be upper bounded by $t_p$, such that minimizing $\sum_p t_p$ under the constraint
\begin{align}
    \textbf{C5:}\quad\sum_i t_p v_{i,p} z_{i,p} \geq 1
\end{align}

\noindent is equivalent to minimizing the original objective. The resulting bilinear constraint including a multiplication between a continuous variable $t_p$ and a binary variable $v_{i,p}$ is still non-convex. However, using Big-M formulation \cite{Wolsey_Integer_2021}, this bilinear term can be linearized by introducing a sufficiently large constant $t_{p,\max}$, chosen empirically to upper-bound the worst case CRLB. Let $t'_{i,p} = t_p v_{i,p}$, then the original bilinear constraint can be equivalently expressed by the set of linear constraints
\begin{align}
    \textbf{C6}\quad\textbf{a:}\quad&0 \leq t_p \leq t_{p,\max}, \label{eq:BigM1}\\
    \textbf{b:}\quad&0 \leq t'_{i,p} \leq t_{p,\max} v_{i,p}, \label{eq:BigM2}\\
    \textbf{c:}\quad&t'_{i,p} \leq t_p, \label{eq:BigM3}\\
    \textbf{d:}\quad&t'_{i,p} \geq t_p - t_{p,\max}(1 - v_{i,p}). \label{eq:BigM4}
\end{align}
\subsubsection{Binary Relaxation and Penalization}
While some solvers, i.e., MOSEK \cite{mosek}, can handle integer variables directly, the large number of binary variables in \textbf{C4} makes the problem computationally challenging. We therefore adopt a hybrid approach, relaxing the sensor selection variables $v_{i,p}$ and the fronthaul allocation variables $a_{i,j}^{(k)}$ to continuous values on the interval $[0,1]$. The backhaul allocation variables $b_j^{(k)}$, however, are retained as binary. This is motivated by two factors: first, the number of backhaul variables is significantly smaller than the number of sensing and fronthaul variables, making them amenable to direct integer optimization. Second, retaining their binary nature prevents an overly loose relaxation, which could lead to suboptimal solutions.

To penalize fractional relaxed solutions and promote binary decisions, we introduce the regularization term
\begin{align}
    \label{eq:G_reg}
    % G_\text{reg} &= \lambda_v \sum_{i,p} v_{i,p}\left(1-v_{i,p}\right) + \lambda_a \sum_{i,j,k} a_{i,j}^{(k)}\left(1-a_{i,j}^{(k)}\right)\nonumber\\& \quad+ \lambda_b \sum_{j,k} b_{j}^{(k)}\left(1-b_{j}^{(k)}\right),
    G_\text{reg} \hspace{-0.05cm}&= \hspace{-0.05cm}\sum_{i,p} \hspace{-0.1cm}\lambda_{i,p} v_{i,p}\left(1-v_{i,p}\right) + \sum_{i,j,k} \hspace{-0.1cm}\eta^{(k)}_{i,j} a_{i,j}^{(k)}\left(1-a_{i,j}^{(k)}\right),
    \vspace{-0.3cm}
\end{align}
where $\lambda_{i,p}, \eta^{(k)}_{i,j} > 0, \forall i,j,k,p$ are regularization parameters for each individual link.
The quadratic terms in $G_{\text{reg}}$ are concave, making the overall objective function a difference of convex functions, which cannot be solved directly. 
\subsubsection{Successive Convex Approximation}
The structure within \eqref{eq:G_reg} allows us to apply SCA method by which the concave penalty is replaced by its first-order Taylor expansion and linearized around the current solution. The Problem is then solved iteratively, updating the linearization point in each iteration.
Let the solution from the previous iteration be denoted by $v_{i,p}^{(*)}, a_{i,j}^{(k,*)}, \forall i,j,p,k$.
The resulting optimization problem for each iteration is then given by
\begin{mini!}|s|
    {\substack{v_{i,p},a_{i,j}^{(k)}, b_{j}^{(k)}, \\ t_p, t'_{i,p}, \forall i,j,p,k}}{\sum_p t_p + G_{\text{lin}}^{(*)}}%
    {\label{Eq:P3n}}{\textbf{P2:}}
    \addConstraint{\hspace{-1cm}\textbf{C1}, \textbf{C2}, \textbf{C3}, \textbf{C6}}{\nonumber}
    \addConstraint{\hspace{-1cm} \widetilde{\textbf{C4:}}\;v_{i,p}, a_{i,j}^{(k)}, \in (0,1),}{\;\forall i,j,p,k}
    \addConstraint{\hspace{-1cm} b_j^{(k)} \in \{0,1\},}{\;\forall j,k.}
\end{mini!}
where the linearized regularization term is
\begin{align}
    % G_{\text{lin}}^{(*)} = \sum_{i,p} \lambda_v v_{i,p} \left( 1 - 2 v_{i,p}^{(*)} \right)& + \sum_{i,j,k} \lambda_a a_{i,j}^{(k)} \left( 1 - 2 a_{i,j}^{(k,*)} \right) \nonumber \\
    % + \sum_{j,k}& \lambda_b b_{j}^{(k)} \left( 1 - 2 b_{j}^{(k,*)} \right).
    G_{\text{lin}}^{(*)} \hspace{-0.1cm}= \hspace{-0.1cm}\sum_{i,p}\hspace{-0.1cm} \lambda_{i,p} v_{i,p} \hspace{-0.1cm}\left( 1 \hspace{-0.05cm}- \hspace{-0.05cm}2 v_{i,p}^{(*)} \right)& \hspace{-0.1cm}+ \hspace{-0.1cm}\sum_{i,j,k}\hspace{-0.1cm} \eta^{(k)}_{i,j} a_{i,j}^{(k)} \hspace{-0.1cm}\left( 1\hspace{-0.05cm}-\hspace{-0.05cm}2 a_{i,j}^{(k,*)} \right)\hspace{-0.1cm}.
\end{align}
\subsection{Proposed Algorithm \& Complexity Analysis}
The algorithm proceeds in two nested loops. The inner loop solves the MILP defined in \textbf{P2}, where the concave penalty terms for the relaxed variables $v_{i,p}$ and $a_{i,j}^{(k)}$ are linearized around the current iterate. The outer loop adaptively updates the penalty parameters $\lambda_{i,p}$ and $\eta^{(k)}_{i,j}$ based on the binary violation of their respective variables. If the violation for a variable exceeds a predefined threshold, its corresponding penalty is increased to steer it toward an integer value in subsequent iterations. This process continues until all relaxed variables have converged to be sufficiently close to binary.
Upon convergence, the solution to the relaxed problem is projected back to the integer domain by rounding the relaxed variables to the nearest integer.
The complete procedure is detailed in Algorithm~\ref{alg:sca_solver}.

\begin{algorithm}[t]
\caption{SCA-MILP Resource Allocation}\label{alg:sca_solver}
\resizebox{0.8\linewidth}{!}{%
\begin{minipage}{\linewidth}
\DontPrintSemicolon
\SetKwInOut{Input}{Input}\SetKwInOut{Output}{Output}
\Input{$z_{i,p}$, Capacities, Max iterations $Q_\text{max}$}
\Output{$v^*, a^*, b^*$}
Initialize $v^{(0)}, a^{(0)}$; Set penalties $\lambda_{i,p}, \eta_{i,j,k} > 0$.\;
\For{$n_\text{out} = 1, \dots, Q_\text{max}$}{
    \Repeat{converged}{
        Solve \textbf{P2} as MILP \;
        Update reference points $v^{(*)} \gets v$, $a^{(*)} \gets a$.\;
    }
    \uIf{$\exists_{i,p}: |v_{i,p} - \lfloor v_{i,p} \rceil| > 0.1$}{
        Update all $\lambda_{i,p}$ where violation $> 0.1$.\;
    }
    \uIf{$\exists_{i,j,k}: |a_{i,j}^{(k)} - \lfloor a_{i,j}^{(k)} \rceil| > 0.1$}{
        Update all $\eta_{i,j,k}$ where violation $> 0.1$.\;
    }
    \Else{
        \textbf{break}
    }
}
\Return{$v^* \gets \lfloor v \rceil$, $a^* \gets \lfloor a \rceil$, $b^* \gets b$}
\end{minipage}%
}
\end{algorithm}

The computational complexity of the proposed algorithm is dominated by the iterative solution of the convex subproblem \textbf{P2}. 
Since the backhaul allocation variables are kept binary, the complexity of solving an MILP is generally exponential in the number of binary variables. It is typically solved using a branch-and-bound algorithm \cite{mosek}, where each node requires the solution of a linear program which is solvable in polynomial time using interior point methods. Specifically, the computational complexity per iteration and per branch is $\mathcal{O}((N_{\text{var}} + N_{\text{const}})^{3})$, where $N_{\text{var}}$ and $N_{\text{const}}$ represent the total number of decision variables and constraints, respectively. Given that $N_{\text{var}}=2IP+KIJ+KJ+P$ and $N_{\text{const}}=I+J+2K+3IP+KIJ+KJ$, the complexity is dominated by $\mathcal{O}\left((KIJ)^{3}+(PI)^{3}\right)$ per branch and iteration. The largest impact on complexity arises from the number of sensors $I$ as it contributes in both terms. The overall complexity additionally scales with the number of binary variables ($JK$) and the number of SCA iterations. 

In practice, the computationally demanding allocation update only needs to be performed when the sensing geometry or network conditions change significantly, allowing the resulting allocation scheme to be reused across multiple buffered measurement batches. 

\section{Simulation Results and Discussion}

For the simulation we consider a cloud radar network consisting of two ES and one FC. Sensors are uniformly distributed in a circle with radius $10\,$m containing four randomly placed targets. 
The radar sensors with $M=8$ antenna elements collect measurements using $L=256$ chirps within a frame duration of $T_F=50\,$ms and $N=256$ samples quantized with $12$ bits and a noise figure of $10$\,dB. The fronthaul and backhaul links operate at carrier frequency $f_c = 1.9\,$GHz with bandwidth $B_c = 1.728\,$MHz divided into $20$ subcarriers. We utilize $10$ of them for the fronthaul and the remaining for the backhaul links. The frame time $T_F$ is divided into $10$ timeslots, resulting in $K=100$ time-frequency resource blocks for both fronthaul and backhaul phases. The window size for spectral compression is set to $\Delta_r = 32$, $\Delta_d = 32$, and $\Delta_a = 4$.

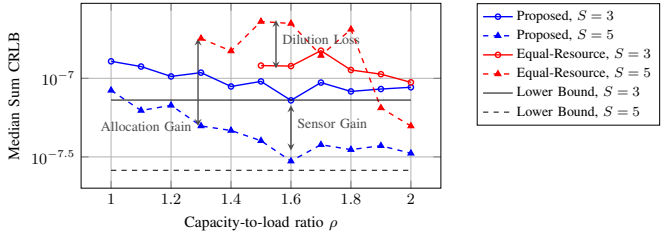
\begin{figure}
    \centering
    \resizebox{\linewidth}{!}{
    \begin{tikzpicture}
\begin{axis}[
    width=\linewidth,
    height=0.6\linewidth,
    xlabel={Capacity-to-load ratio $\rho$},
    ylabel={Median Sum CRLB},
    ymode=log,
    ymin=2e-8, ymax=3e-7,
    grid=both,
    grid style={line width=0.1pt, draw=gray!30},
    major grid style={line width=0.2pt, draw=gray!60},
    legend style={at={(1.1,1.0)}, anchor=north west,
                  font=\footnotesize, cells={anchor=west}},
    legend columns=1,
    tick label style={font=\small},
    label style={font=\small},
    mark options={solid},
]

\addplot [solid, blue, thick, mark=o, mark size=1.5pt] coordinates {
    (1.0, 1.282922e-07)
    (1.1, 1.188107e-07)
    (1.2, 1.028032e-07)
    (1.3, 1.086591e-07)
    (1.4, 8.873842e-08)
    (1.5, 9.556881e-08)
    (1.6, 7.240385e-08)
    (1.7, 9.393426e-08)
    (1.8, 8.252920e-08)
    (1.9, 8.545782e-08)
    (2.0, 8.771265e-08)
};
\addlegendentry{Proposed, $S=3$}

\addplot [dashed, blue, thick, mark=triangle*, mark size=1.8pt] coordinates {
    (1.0, 8.384171e-08)
    (1.1, 6.242214e-08)
    (1.2, 6.730261e-08)
    (1.3, 4.975034e-08)
    (1.4, 4.657370e-08)
    (1.5, 4.007570e-08)
    (1.6, 2.972322e-08)
    (1.7, 3.780565e-08)
    (1.8, 3.519157e-08)
    (1.9, 3.724520e-08)
    (2.0, 3.333993e-08)
};
\addlegendentry{Proposed, $S=5$}

\addplot [solid, red, thick, mark=o, mark size=1.5pt] coordinates {
    (1.5, 1.204969e-07)
    (1.6, 1.196380e-07)
    (1.7, 1.499662e-07)
    (1.8, 1.129751e-07)
    (1.9, 1.062708e-07)
    (2.0, 9.426278e-08)
};
\addlegendentry{Equal-Resource, $S=3$}

\addplot [dashed, red, thick, mark=triangle*, mark size=1.8pt] coordinates {
    (1.3, 1.793183e-07)
    (1.4, 1.492582e-07)
    (1.5, 2.298206e-07)
    (1.6, 2.233046e-07)
    (1.7, 1.396833e-07)
    (1.8, 2.050713e-07)
    (1.9, 6.485273e-08)
    (2.0, 4.970936e-08)
};
\addlegendentry{Equal-Resource, $S=5$}

\addplot [solid, black!99, thin] 
    coordinates {(1.0, 7.272368e-08) (2.0, 7.272368e-08)};
\addlegendentry{Lower Bound, $S=3$}

\addplot [dashed, black!99, thin] 
    coordinates {(1.0, 2.599227e-08) (2.0, 2.599227e-08)};
\addlegendentry{Lower Bound, $S=5$}

\draw[stealth-stealth, thick, black!70]
    (axis cs:1.29,1.793183e-07)
    -- (axis cs:1.29,4.975034e-08)
    node[rotate=0, left, pos=1, font=\footnotesize] {Allocation Gain};

\draw[stealth-stealth, thick, black!70]
    (axis cs:1.6,6.740385e-08)
    -- (axis cs:1.6,3.472322e-08)
    node[rotate=0, right, pos=0.4, font=\footnotesize] {Sensor Gain};

\draw[stealth-stealth, thick, black!70]
    (axis cs:1.55,1.154969e-07)
    -- (axis cs:1.55,2.348206e-07)
    node[rotate=0, right, pos=0.6, font=\footnotesize] {Dilution Loss};

\end{axis}
\end{tikzpicture}}
    \caption{Median of the sum CRLB for the proposed method and the baseline as a function of the capacity-to-load ratio.}
    \label{fig:Results}
\end{figure}

We compare the performance of the proposed dynamic resource allocation method against a static baseline that allocates resources equally across all sensors. The results, presented in Fig.~\ref{fig:Results}, show the sum of the normalized CRLBs for all targets as a function of the network capacity, normalized by the total sensor flow rate. The data is derived from $50$ Monte Carlo trials with randomized target placements and channel realizations.

The baseline's performance degrades as the network capacity becomes more constrained and eventually breaks down when the capacity per sensor becomes insufficient to cover all targets. In contrast, the proposed method maintains a reasonable performance even at low communication capacities by prioritizing the most informative sensors and allocating resources accordingly. Both methods converge in the high-capacity regime, i.e. $\rho > 2$, to the maximum achievable estimation accuracy given by the aggregated data across \textit{all sensors}, indicated by the black solid and dashed horizontal lines in Fig.~\ref{fig:Results}.

Furthermore, the addition of more sensors highlights a key difference between the two approaches. In our proposed framework, more sensors improve sensing performance by providing a larger selection of informative views (Sensor Gain). For the baseline, however, adding sensors dilutes the fixed resource allocation, leading even to degraded performance (Dilution Loss). 

\section{Conclusion}
This paper investigated the resource allocation problem for distributed radar sensing in capacity-constrained cloud networks. We proposed a buffered access protocol with local data compression and formulated the joint sensor selection and resource allocation as a mixed-integer problem to minimize the aggregate CRLB. A solution based on Big-M formulation and SCA was developed. Simulation results demonstrated that our method significantly outperforms a baseline with static resource allocation, especially under tight capacity constraints, by dynamically prioritizing the most informative sensors. Future work will explore more complex network topologies and moving target scenarios.

% references section

% can use a bibliography generated by BibTeX as a .bbl file
% BibTeX documentation can be easily obtained at:
% http://mirror.ctan.org/biblio/bibtex/contrib/doc/
% The IEEEtran BibTeX style support page is at:
% http://www.michaelshell.org/tex/ieeetran/bibtex/
%\bibliographystyle{IEEEtran}
% argument is your BibTeX string definitions and bibliography database(s)
%\bibliography{IEEEabrv,../bib/paper}
%
% <OR> manually copy in the resultant .bbl file
% set second argument of \begin to the number of references
% (used to reserve space for the reference number labels box)

\bibliographystyle{IEEEtran}
\bibliography{IEEEabrv,references}

% \begin{thebibliography}{1}

% \bibitem{IEEEhowto:kopka}
% H.~Kopka and P.~W. Daly, \emph{A Guide to \LaTeX}, 3rd~ed.\hskip 1em plus
%   0.5em minus 0.4em\relax Harlow, England: Addison-Wesley, 1999.

% \end{thebibliography}

% that's all folks
\end{document}